\documentclass[11pt]{article}

\usepackage{graphicx}
\usepackage{amssymb}
\usepackage{amsmath}
\usepackage{upgreek}
\usepackage{multirow}
\usepackage[latin1]{inputenc}
\usepackage[labelfont=bf,labelsep=period,footnote size]{caption}
\usepackage{color}
\usepackage[breaklinks=true,colorlinks=true,linkcolor=blue,citecolor=blue,backref=none,pagebackref=false]{hyperref}
\usepackage{setspace}
\usepackage{epstopdf}

\numberwithin{equation}{section}

\pdfoutput=1
\usepackage{geometry}
\usepackage[auth-sc-lg,affil-sl]{authblk}

\makeatletter
\renewcommand{\@biblabel}[1]{\quad#1.} 
\makeatother

\usepackage{setspace}
\newcommand{\keywords}[1]{\begin{center}\textbf{Keywords:} #1\end{center}}

\newcommand{\fbar}{\overline{f}}
\newcommand{\eff}{\rm eff}
\newcommand{\matr}[1]{\underline{\underline{#1}}}
\newcommand{\vect}[1]{\underline{#1}}
\newcommand{\idmatr}{\mathbb{I}}

\begin{document}

\title{Fixation in finite populations evolving in fluctuating environments}

\author[1]{Peter Ashcroft \thanks{peter.ashcroft@postgrad.manchester.ac.uk}}
\affil[1]{Theoretical Physics, School of Physics and Astronomy, The University of Manchester, Manchester M13 9PL, United Kingdom}

\author[2]{Philipp M.~Altrock \thanks{paltrock@jimmy.harvard.edu}}
\affil[2]{Program for Evolutionary Dynamics, Harvard University, Cambridge, MA 02138, USA}
\affil[2]{Harvard School of Public Health, Boston, MA 02115, USA}
\affil[2]{Dana-Farber Cancer Institute, Boston, MA 02215, USA}

\author[1]{Tobias Galla \thanks{tobias.galla@manchester.ac.uk} \thanks{Corresponding author}}

\label{firstpage}

\maketitle

\begin{abstract}
The environment in which a population evolves can have a crucial impact on selection. We study evolutionary dynamics in finite populations of fixed size in a changing environment. The population dynamics are driven by birth and death events. The rates of these events may vary in time depending on the state of the environment, which follows an independent Markov process. We develop a general theory for the fixation probability of a mutant in a population of wild-types, and for mean unconditional and conditional fixation times. We apply our theory to evolutionary games for which the payoff structure varies in time. The mutant can exploit the environmental noise; a dynamic environment that switches between two states can lead to a probability of fixation that is higher than in any of the individual environmental states. We provide an intuitive interpretation of this surprising effect. We also investigate stationary distributions when mutations are present in the dynamics. In this regime, we find two approximations of the stationary measure. One works well for rapid switching, the other for slowly fluctuating environments.
\end{abstract}
\bigskip
\newpage

\onehalfspacing

\section{Introduction}
Evolutionary dynamics describes the change of populations over time subject to spontaneous mutation, selection, and other random events \cite{vincent:book:2005,nowak:book:2006}. Different phenotypes in the population can emerge spontaneously by mutation, i.e.~through errors during reproduction of wild-types. In many cases, wild-type and mutant individuals are characterised by heritable differences in behavioural traits or strategies \cite{nowak:book:2006}. Selection acts on different phenotypes and changes the population composition. Changes in the state of the environment can alter these selective pressures over time.

Time-varying environments are relevant in the evolution of bacterial populations subject to environment modulations by a host \cite{franzenburg:ISME:2013,mcFall-Ngai:PNAS:2013}, or varying antibiotic stress. An illustrative example is the evolution of normal (wild-type) cells and resistant `persister' cells (mutant). This was examined by Kussell {\em et al.~}\cite{kussell:Genetics:2005}, where periods of antibiosis were turned on and off. During times of antibiotic stress the growth rate of normal cells was reduced, but the resistant cells sustain population levels. In addition, Acar {\em et al.~}\cite{acar:NG:2008} provided further experimental evidence supporting the deterministic model used in \cite{kussell:Genetics:2005}. More complicated studies of dynamics in switching environments rely on cells `sensing' the environment \cite{kussell:Science:2005} and on the history or information of the environment during a cell's life \cite{leibler:PNAS:2010,rivoire:JSP:2011}. These examples illustrate that the assumption of an interaction structure independent of time is not always realistic. At the same time it is largely an open question how complex interactions between phenotypes together with spontaneous changes in the environment influence the evolutionary dynamics. 

The interactions of phenotypes in a population can be formalised in an evolutionary game \cite{hofbauer:book:1998,gintis:book:2009}. Such games can be used to describe conflict over food or territory, cheating in resource allocation, as well as interactions between variants of a gene \cite{maynard-smith:Nature:1973,milinski:Nature:1987,maclean:PloSB:2010,traulsen:JTB:2012,altrock:PlosCB:2011}. In an evolutionary normal-form game each individual can be associated with one out of a finite set of strategies. A payoff matrix quantifies the reward received by a given individual when it interacts with another individual \cite{gintis:book:2009}.

The dynamics of populations interacting in such a game are often described by deterministic replicator equations or similar differential equations \cite{hofbauer:book:1998,taylor:MB:1978,sandholm:book:2010}. While deterministic dynamics are useful to understand the action of selection \emph{per se}, more interesting phenomena arise when stochastic effects are taken into account. A stochastic approach is appropriate~--~often even strictly required~--~to understand the impact of fluctuations in finite populations \cite{lenormand:TREE:2009,black:TREE:2012}. Deterministic approaches fail to capture effects such as fixation and extinction, or the convergence to a stationary distribution in systems with mutation \cite{antal:BMB:2006,black:PRL:2012,altrock:JTB:2012,arnoldt:JRSI:2012,du:JRSI:2014}.

There is an increasing body of literature on stochastic evolutionary games. For example, analytical results for the probabilities of a single mutant to reach fixation have been obtained \cite{lessard:JMB:2007,szabo:PR:2007,traulsen:bookchapter:2009,perc:BioSys:2010}. However, most of this existing work focusses on games played in a fixed environment; the underlying payoff matrix itself remains unchanged in time. This assumption may not be appropriate in cases where external factors influence the environment.

External fluctuations in evolutionary games have been previously introduced by adding extrinsic noise to continuous model parameters \cite{assaf:PRL:2013}, or by letting strategy space itself vary in time \cite{huang:NatComm:2012}. Environmental variability has also been the subject of investigation in predator-prey models \cite{dobramysl:PRL:2013,dobramysl:JStatMech:2013}.

In this article we explore different theoretical approaches that allow calculations of fixation probabilities and mean fixation times of a rare mutation, under fluctuating environmental conditions. We use a generic birth-death framework, as described in section~\ref{sec:model}, and our results thus apply to a wide class of population dynamics. In section~\ref{sec:gen} the theory is developed for an environment which can transition between an arbitrary number of discrete states, and we expand on the two-environment scenario in section~\ref{sec:twostate}. To illustrate our theoretical results we study the fixation properties in an evolutionary game that stochastically switches between a coexistence game and a coordination game in section~\ref{sec:games}. We determine environmental conditions under which the success of a rare invading mutant is maximal. This is seen to occur at a non-trivial combination of switching rates.
For the case in which mutations occur during the dynamics, as described in section~\ref{sec:mut}, we explore how the stationary distribution of the population changes in fluctuating environments. We derive approximations for the stationary distribution, valid for a large range of switching rates. We summarise our findings in section~\ref{sec:sum} and put them into context.

\section{Mathematical Model}\label{sec:model}
\begin{figure}[t]
\begin{center} \includegraphics[width=0.75\textwidth]{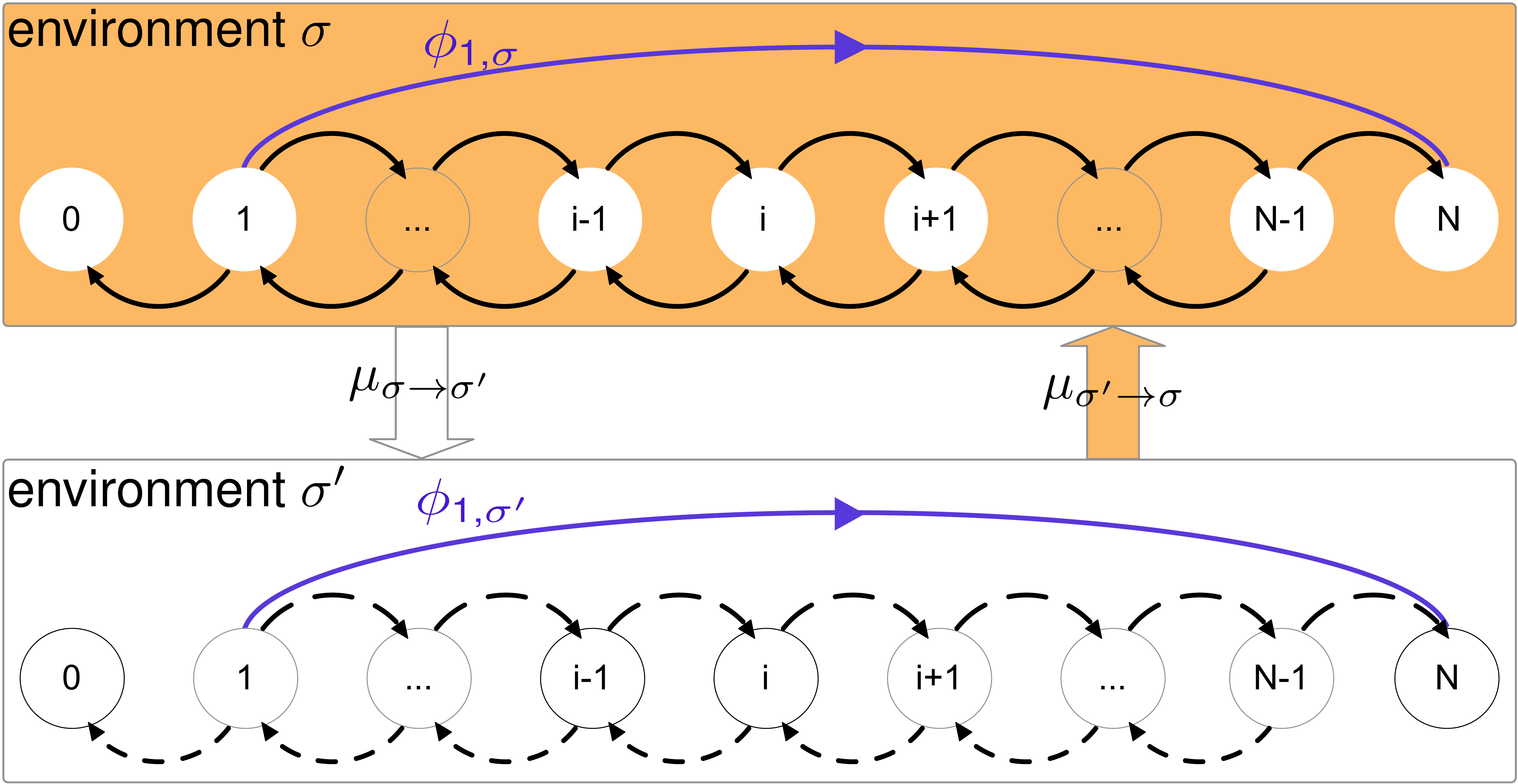}
\caption{
A population undergoes a one-step birth-death process, such that given the population is in state $i$, in one time-step it may transition to $i-1$ or $i+1$, or remain at $i$. The states $i=0$ and $i=N$ are absorbing in both environments (no arrows out of these states). The transition probabilities (birth/death rates) are dependent on the state of the environment, indicated by solid vs. dashed arrows in environments $\sigma$ and $\sigma'$, respectively. The environment switches from state $\sigma$ to $\sigma'$ with probability $\mu_{\sigma\to\sigma'}$ in any one time-step. The quantity $\phi_{i,\sigma}$ represents the probability of fixation, as discussed in section \ref{sec:gen}.
(Online version in colour.)
}
\label{fig:fig1}
\end{center}
\end{figure}

We seek to model evolutionary dynamics in finite populations of two species that are subject to environmental changes. The changes in the environment are such that at any given point in time, the system can be in one of a finite set of environmental states. The state that the environment is in determines the details of the birth and death dynamics. We focus on two cases: 
first, in the absence of mutations in the dynamics, we derive laws to predict the probability and mean time of the fixation of a mutant. Fixation describes the event in which mutants take over the population as opposed to going extinct. Fixation or extinction are the only two outcomes of dynamics of a rare mutant in a finite population \cite{ewens:book:2004}. In figure~\ref{fig:fig1} we show a basic sketch of this scenario in which the two monomorphic states of the population are absorbing. Second, we study the case when mutations occur in the dynamics. There are then no absorbing states. Instead, the dynamics converges to a stationary distribution.

\subsection{Birth-death dynamics}
We consider populations consisting of a fixed number of $N$ individuals. Each individual can be of one of two types, $A$ or $B$, which we refer to as `mutant' and `wild-type', respectively. The population is well mixed; every individual can interact with any other individual. The state of the population is fully characterised by the number, $i$, of individuals of type $A$. The remaining $N-i$ individuals are of type $B$. We furthermore assume that at any one time the environment can be in one of $\Omega$ discrete states, labelled $\sigma\in\Lambda$, where $\Lambda$ is the space of states of the environment ($|\Lambda|=\Omega$). Hence the state of the entire system at any time is given by the pair $(i,\sigma)$.

The discrete-time birth-death dynamics of the population for a given environment, $\sigma$, is then specified by the transition probabilities $\omega_{i,\sigma}^+$ and $\omega_{i,\sigma}^-$ of a one-step process. Specifically, if the system is in state $(i,\sigma)$ the population transitions to state $i+1$ in the next time step with probability $\omega_{i,\sigma}^+$. Similarly the state of the population in the next time step is $i-1$ with probability $\omega_{i,\sigma}^-$. These transitions are shown as black arrows in figure~\ref{fig:fig1}. With probability $1-(\omega_{i,\sigma}^++\omega_{i,\sigma}^-)$ the population remains in state $i$. We always assume that $\omega_{i,\sigma}^\pm\geq 0$ and $\omega_{i,\sigma}^++\omega_{i,\sigma}^-\leq 1$ for all $(i,\sigma)$.

With the exception of section~\ref{sec:mut} we will always assume that the states $i=0$ (all-$B$) and $i=N$ (all-$A$) are absorbing ($\omega_{0,\sigma}^+=0$ and $\omega_{N,\sigma}^-=0$ for all $\sigma\in\Lambda$). In the absence of further mutation events a type, once absent, can never be re-introduced. If mutations are present in the dynamics, then the states $i=0$ and $i=N$ are no longer absorbing and the system converges to a unique, non-trivial stationary state. We consider this case in section~\ref{sec:mut}. 

\subsection{Fluctuating environment \label{ssec:switchingenv}}
In our approach the environment evolves from one state to another {\em independently} of the state of the population. This simplification still captures a wide array of natural scenarios. In the discrete-time setup we take the dynamics of the environment as a simple Markov chain, described by the transition matrix $\matr{\mu}=(\mu_{\sigma\to\sigma'})$ of size $\Omega\times\Omega$. The entry $\mu_{\sigma\to\sigma'}$ represents the probability that the environment changes to state $\sigma'$ in the next time-step, if it is currently in state $\sigma$, as shown in figure~\ref{fig:fig1}. The matrix $\matr{\mu}$ is a stochastic matrix, $\sum_{\sigma'}\mu_{\sigma\to\sigma'}=1$ for all $\sigma\in\Lambda$. 

A switch of the environment effectively modifies the birth-death dynamics in the population. We do not specify the exact type of interaction at this point, but keep the rates $\omega_{i,\sigma}^\pm$ general.

\section{Fixation probability and time for a general birth-death process in a fluctuating environment}\label{sec:gen}

\subsection{Fixation probability}\label{ssec:discreteFP}
Let us first consider a discrete-time evolutionary process. If the system is in state $(i,\sigma)$ at a given time, it may transition to $3\Omega$ possible states in any one time-step. These are given by $(i,\sigma')$, $(i+1,\sigma')$ and $(i-1,\sigma')$, where $\sigma'\in\Lambda$ can be any of the $\Omega$ states of the environment. If we write $R_{(i,\sigma)\to(j,\sigma')}$ for the probability of a transition from $(i,\sigma)$ to $(j,\sigma')$, we have
\begin{eqnarray}
R_{(i,\sigma)\to(i+1,\sigma')}&=&\mu_{\sigma\to\sigma'}\omega_{i,\sigma}^+ \nonumber \\
R_{(i,\sigma)\to(i-1,\sigma')}&=&\mu_{\sigma\to\sigma'}\omega_{i,\sigma}^- \nonumber \\
R_{(i,\sigma)\to(i,\sigma')}~~&=&\mu_{\sigma\to\sigma'}(1-\omega_{i,\sigma}^+-\omega_{i,\sigma}^-).
\label{eq:reactionprobs}
\end{eqnarray}
No transitions from $(i,\sigma)$ to $(j,\sigma')$ can occur when $|i-j|>1$.  In this setup the birth-death probabilities are determined by the state of the environment at the beginning of the discrete time-step.

The fixation probability, $\phi_{i,\sigma}$, is the probability that the system ends up in the absorbing state with $N$ individuals of type $A$, conditioned on initial state $(i,\sigma)$. The probability of fixation of a single mutant, $\phi_{1,\sigma}$, is of particular interest \cite{antal:BMB:2006}. It is briefly illustrated in figure~\ref{fig:fig1}. In our scenario with switching between $\Omega$ environmental states, following the lines of Refs.~\cite{nowak:book:2006,traulsen:bookchapter:2009,karlin:book:1981}, the following balance equation for the fixation probabilities can be found 
\begin{equation}
\phi_{i,\sigma} = \sum_{\sigma' \in \Lambda} \mu_{\sigma\to\sigma'}\left[\omega^+_{i,\sigma}\phi_{i+1,\sigma'}+ \omega^-_{i,\sigma}\phi_{i-1,\sigma'} +(1-\omega^+_{i,\sigma}-\omega^-_{i,\sigma}) \phi_{i,\sigma'}\right].
\label{eq:fixprobbalance}
\end{equation}
This is to be solved along with the boundary conditions $\phi_{0,\sigma}=0$ and $\phi_{N,\sigma}=1$ for all $\sigma \in \Lambda$.

To obtain a formal solution, we introduce $\psi_{i,\sigma}=\sum_{\sigma'}\mu_{\sigma\to\sigma'}\phi_{i,\sigma'}$, or in matrix form $\vect{\psi}_i=\matr{\mu}\cdot\vect{\phi}_i$. The vectors $\vect{\psi}_i$ and $\vect{\phi}_i$ each have $\Omega$ components. The boundary conditions $\phi_{0,\sigma}=0$ and $\phi_{N,\sigma}=1$ translate into $\psi_{0,\sigma}=0$ and $\psi_{N,\sigma}=1$ for all $\sigma\in\Lambda$.
With this notation we have
\begin{equation}
\phi_{i,\sigma} = \omega_{i,\sigma}^+(\psi_{i+1,\sigma}-\psi_{i,\sigma})-\omega_{i,\sigma}^-(\psi_{i,\sigma}-\psi_{i-1,\sigma})+\psi_{i,\sigma}.
\label{eq:fixprobbalance2}
\end{equation}
Using $\vect{\phi}_i=\matr{\mu}^{-1}\cdot\vect{\psi}_i$, we obtain
\begin{equation}
\psi_{i+1,\sigma}-\psi_{i,\sigma} = \gamma_{i,\sigma}\left(\psi_{i,\sigma}-\psi_{i-1,\sigma}\right)+\frac{1}{\omega_{i,\sigma}^+} \left[(\matr{\mu}^{-1}\cdot\vect{\psi}_i)_{\sigma}-\psi_{i,\sigma}\right],
\label{eq:fixprobbalance3}
\end{equation}
where $\gamma_{i,\sigma}=\omega_{i,\sigma}^-/\omega_{i,\sigma}^+$. We stress that the calculation of fixation probabilities and mean fixation times using this formalism requires the matrix $\matr{\mu}$ to be invertible. We comment on this further in the context of a specific example in Sec. \ref{sec:twostate}.

To keep the notation compact we define the variable $\upsilon_{i,\sigma}=\psi_{i,\sigma}-\psi_{i-1,\sigma}$. Using $\psi_{0,\sigma}=0$, we have $\psi_{i,\sigma}=\sum_{j=1}^i \upsilon_{j,\sigma}$. With this notation we can write equation~(\ref{eq:fixprobbalance3}) in the following form
\begin{equation}
\upsilon_{i+1,\sigma} = \gamma_{i,\sigma}\upsilon_{i,\sigma}+\frac{1}{\omega^+_{i,\sigma}}\left[\left(\matr{\mu}^{-1}-\idmatr\right) \cdot \sum_{j=1}^i \vect{\upsilon}_j \right]_\sigma,
\label{eq:fixprobbalance4}
\end{equation}
where $\idmatr$ is the $\Omega\times\Omega$ identity matrix.
This relation expresses the vector $\vect{\upsilon}_{i+1}$ in terms of the vectors $\vect{\upsilon}_1, \vect{\upsilon}_2,\dots, \vect{\upsilon}_i$. We can therefore express all vectors $\vect{\upsilon}_i$ (i=$2,\dots,N$) in terms of $\vect{\upsilon}_1$. The constraint $\sum_{i=1}^N \vect{\upsilon}_i = \vect{\psi}_N = (1,\dots,1)^{\rm T}$ then determines $\vect{\upsilon}_1$ self-consistently. We note that the resulting set of equations is linear in the set $\{\upsilon_{1,\sigma}\}$. Hence a solution can be obtained in closed form, in principle. In practice one inverts the linear system using one of the standard algebraic manipulation packages. Once $\vect{\upsilon}_1$ has been found, the other components $\vect{\upsilon}_i$, with $i=2,\dots,N$, can be computed via equation~(\ref{eq:fixprobbalance4}). One then uses $\vect{\phi}_i=\matr{\mu}^{-1}\cdot\sum_{j=1}^i \vect{\upsilon}_j$ to find the fixation probabilities starting with $i$ individuals of type $A$ in environment $\sigma$, $\{\phi_{i,\sigma}\}$.

We note here that algebraically inverting the linear system~(\ref{eq:fixprobbalance4}) when $N$ is large is difficult due to the very large number of terms in the corresponding expressions. Thus, at present, this theory is limited computationally to relatively small $N$.

In the case of a single environment, $\Omega=1$, the matrix $\matr{\mu}$ is simply the $1\times 1$ identity matrix, and equation~(\ref{eq:fixprobbalance4}) simplifies to the well-known result for discrete-time birth-death processes \cite{nowak:book:2006,traulsen:bookchapter:2009,karlin:book:1981}.

\subsection{Unconditional fixation time}
We write $t_{i,\sigma}$ for the expected number of time-steps taken to reach any one of the two absorbing states, given that the system is started in state $(i,\sigma)$. These fulfil the boundary conditions $t_{0,\sigma}=t_{N,\sigma}=0$. With these definitions we find the following relation
\begin{equation}
t_{i,\sigma} = \sum_{\sigma' \in \Lambda} \mu_{\sigma\to\sigma'}\left[\omega^+_{i,\sigma}t_{i+1,\sigma'}+ \omega^-_{i,\sigma}t_{i-1,\sigma'} +(1-\omega^+_{i,\sigma}-\omega^-_{i,\sigma}) t_{i,\sigma'}\right] + 1.
\label{eq:uncondbalance1}
\end{equation}
Introducing the variable $\xi_{i,\sigma}=\sum_{\sigma'}\mu_{\sigma\to\sigma'}t_{i,\sigma'}$, we have
\begin{equation}
t_{i,\sigma} = \omega^+_{i,\sigma}(\xi_{i+1,\sigma}-\xi_{i,\sigma}) -\omega^-_{i,\sigma}(\xi_{i,\sigma}-\xi_{i-1,\sigma})+\xi_{i,\sigma}+1,
\label{eq:uncondbalance2}
\end{equation}
and with the notation $\nu_{i,\sigma}=\xi_{i,\sigma}-\xi_{i-1,\sigma}$ we arrive at
\begin{equation}
\nu_{i+1,\sigma}=\gamma_{i,\sigma}\nu_{i,\sigma}+\frac{1}{\omega^+_{i,\sigma}} \left[\left(\matr{\mu}^{-1}-\idmatr\right) \cdot \sum_{j=1}^i \vect{\nu}_j\right]_\sigma -\frac{1}{\omega^+_{i,\sigma}}.
\label{eq:uncondbalance3}
\end{equation}
This relation allows one to express all vectors $\vect{\nu}_i$ ($i=2,\dots,N$) in terms of $\vect{\nu}_1$. The constraint $\sum_{i=1}^N \vect{\nu}_i=(0,\dots,0)^{\rm T}$ then determines $\vect{\nu}_1$, and the mean unconditional fixation times are computed using $\vect{t}_i=\matr{\mu}^{-1}\cdot\sum_{j=1}^i\vect{\nu}_j$.

\subsection{Conditional fixation time}
We write $t_{i,\sigma}^A$ for the mean fixation time conditioned on absorption in the all-$A$ state, given that the system is initially in state $(i,\sigma)$. To find this conditional fixation time, we proceed along similar lines as before. Introducing the variable $\theta_{i,\sigma}=\phi_{i,\sigma}t_{i,\sigma}^A$, which has boundary conditions $\theta_{0,\sigma}=\theta_{N,\sigma}=0$, the following balance equation can be found,
\begin{equation}
\theta_{i,\sigma} = \sum_{\sigma' \in \Lambda} \mu_{\sigma\to\sigma'}\left[\omega^+_{i,\sigma}\theta_{i+1,\sigma'}+ \omega^-_{i,\sigma}\theta_{i-1,\sigma'} +(1-\omega^+_{i,\sigma}-\omega^-_{i,\sigma}) \theta_{i,\sigma'}\right] + \phi_{i,\sigma}.
\label{eq:condbalance1}
\end{equation}
We note that equation (\ref{eq:uncondbalance1}) and equation (\ref{eq:condbalance1}) appear to be very similar, but the difference is more than just a global pre-factor $\phi_{i,\sigma}$; each term in the expression has different indices $i$ and $\sigma$.
  
Introducing the variable $\zeta_{i,\sigma}=\sum_{\sigma'}\mu_{\sigma\to\sigma'}\theta_{i,\sigma'}$, we have
\begin{equation}
\theta_{i,\sigma} = \omega^+_{i,\sigma}(\zeta_{i+1,\sigma}-\zeta_{i,\sigma}) -\omega^-_{i,\sigma}(\zeta_{i,\sigma}-\zeta_{i-1,\sigma})+\zeta_{i,\sigma}+\phi_{i,\sigma},
\label{eq:condbalance2}
\end{equation}
and introducing $\eta_{i,\sigma}=\zeta_{i,\sigma}-\zeta_{i-1,\sigma}$ we arrive at
\begin{equation}
\eta_{i+1,\sigma}=\gamma_{i,\sigma}\eta_{i,\sigma}+\frac{1}{\omega^+_{i,\sigma}}\left[\left(\matr{\mu}^{-1}-\idmatr\right)\cdot\sum_{j=1}^i\vect{\eta}_j\right]_\sigma-\frac{1}{\omega^+_{i,\sigma}}\phi_{i,\sigma}.
\label{eq:condbalance3}
\end{equation}
The set $\{\theta_{i,\sigma}\}$ can then be found using an approach similar to the one described above. Results for the mean conditional fixation time can then be obtained using $t^A_{i,\sigma}=\theta_{i,\sigma}/\phi_{i,\sigma}$.

\subsection{Continuous-time model}
In any of the elementary time steps of the above discrete-time model, {\em both} the state of the population ($i$) and the state of the environment ($\sigma$) can change. We next consider a continuous-time setup. There are two types of discrete events that may occur at any time: (i) the state of the environment  may change, or (ii) a birth-death event may occur. The rate (per unit time) with which a transition from state $\sigma$ to state $\sigma'$ occurs is denoted by $m_{\sigma\to\sigma'}$. The occurrence of these events is independent of the state of the population. The rate with which a birth-death event of the type $i\to i+1$ occurs is $W_{i,\sigma}^+$, if the environment is in state $\sigma$. The rate with which $i\to i-1$ occurs is $W_{i,\sigma}^-$. We write $P_{i,\sigma}(t)$ for the probability to find the system in state $(i,\sigma)$ at time $t$, and find the master equation
\begin{eqnarray}
\partial_t P_{i,\sigma}(t) &=& 
W^+_{i-1,\sigma}P_{i-1,\sigma}(t) -  W^+_{i,\sigma}P_{i,\sigma}(t)
+ W^-_{i+1,\sigma}P_{i+1,\sigma}(t) -  W^-_{i,\sigma}P_{i,\sigma}(t) \nonumber \\
&& +\sum_{\sigma'}\left[m_{\sigma'\to\sigma} P_{i,\sigma'}(t) - m_{\sigma\to\sigma'} P_{i,\sigma}(t)\right].
\label{eq:master}
\end{eqnarray}

\subsubsection*{Fixation probability}
We write $Q_{j,\sigma';i,\sigma}(t)$ for the probability to find the system in state $(j,\sigma')$ a period of time $t$ after it has been started in state $(i,\sigma)$. The corresponding backward master equation \cite{kampen:book:2007,gardiner:book:2009} reads
\begin{eqnarray}
\partial_t Q_{j,\sigma';i,\sigma}(t) &=& 
W^+_{i,\sigma} \bigl[Q_{j,\sigma';i+1,\sigma}(t) - Q_{j,\sigma';i,\sigma}(t)\bigr] 
+ W^-_{i,\sigma} \bigl[Q_{j,\sigma';i-1,\sigma}(t) - Q_{j,\sigma';i,\sigma}(t)\bigr] \nonumber \\
&&+ \sum_{\sigma''}m_{\sigma\to\sigma''}\bigl[Q_{j,\sigma';i,\sigma''}(t)-Q_{j,\sigma';i,\sigma}(t)\bigr].
\label{eq:backwardmaster}
\end{eqnarray}
We define $\varphi_{i,\sigma}(t)=\sum_{\sigma'}Q_{N,\sigma';i,\sigma}(t)$ as the probability that the system has reached fixation in the all-$A$ state a period of time $t$ after the dynamics has been started in state $(i,\sigma)$. This includes fixation before time $t$. By setting $j=N$ and summing over $\sigma'$ in equation (\ref{eq:backwardmaster}) we obtain
\begin{eqnarray}
\partial_t\varphi_{i,\sigma}(t) &=& 
W^+_{i,\sigma}\bigl[\varphi_{i+1,\sigma}(t)-\varphi_{i,\sigma}(t)\bigr] 
+ W^-_{i,\sigma}\bigl[\varphi_{i-1,\sigma}(t)-\varphi_{i,\sigma}(t)\bigr]\nonumber \\
&&+\sum_{\sigma''}m_{\sigma\to\sigma''}\bigl[\varphi_{i,\sigma''}(t)-\varphi_{i,\sigma}(t)\bigr].
\label{eq:backwardmastersum1}
\end{eqnarray}
The fixation probabilities are found as $\phi_{i,\sigma}=\lim_{t\to\infty}\varphi_{i,\sigma}(t)$, and they can be obtained by setting the time derivative in equation~(\ref{eq:backwardmastersum1}) to zero. Introducing $y_{i,\sigma}=\phi_{i,\sigma}-\phi_{i-1,\sigma}$, one finds
\begin{equation}
y_{i+1,\sigma} = g_{i,\sigma}y_{i,\sigma}+ \frac{1}{W_{i,\sigma}^+}\sum_{\sigma'}m_{\sigma\to\sigma'}\left[\sum_{j=1}^i (y_{j,\sigma}-y_{j,\sigma'})\right],
\label{eq:fixprobcont}
\end{equation}
with $g_{i,\sigma}=W_{i,\sigma}^-/W_{i,\sigma}^+$ and where we have used $\phi_{0,\sigma}=0$ to write $\phi_{i,\sigma}=\sum_{j=1}^i y_{j,\sigma}$. For the case of a single environment the second term on the right-hand side vanishes and one recovers again the well known results in single environments \cite{nowak:book:2006,traulsen:bookchapter:2009,goel:book:1974}.
Equation~(\ref{eq:fixprobcont}) has the same general structure as equation~(\ref{eq:fixprobbalance4}). Keeping in mind that $\phi_{N,\sigma}=1$, the fixation probabilities can hence be found by applying the approach outlined in section~\ref{ssec:discreteFP}.

\subsubsection*{Fixation times}
Calculating the mean fixation time using a diffusion approximation \cite{ewens:book:2004,kampen:book:2007,gardiner:book:2009,altrock:PRE:2010} is not appropriate for our model. The environmental switching process has no continuum limit. Instead we work with the backward master equation~(\ref{eq:backwardmaster}) and adapt the calculation outlined by Antal and Scheuring \cite{antal:BMB:2006}.

We introduce $\vartheta_{i,\sigma}(t)=\sum_{\sigma'}\left[Q_{0,\sigma';i,\sigma}(t)+Q_{N,\sigma';i,\sigma}(t)\right]$, the probability that the system has reached fixation in either of the two absorbing states a period of time $t$ after being started in $(i,\sigma)$. Again this includes fixation before $t$. We then have $\rho_{i,\sigma}(t)=\partial_t \vartheta_{i,\sigma}(t)$ for the probability density to reach fixation exactly at time $t$. From the backward master equation (\ref{eq:backwardmaster}) we find
\begin{eqnarray}
\partial_t \rho_{i,\sigma}(t) &=& 
W^+_{i,\sigma} \bigl[\rho_{i+1,\sigma}(t) - \rho_{i,\sigma}(t)\bigr] 
+ W^-_{i,\sigma} \bigl[\rho_{i-1,\sigma}(t)- \rho_{i,\sigma}(t)\bigr] \nonumber \\
&&+ \sum_{\sigma'}m_{\sigma\to\sigma'}\bigl[\rho_{i,\sigma'}(t)-\rho_{i,\sigma}(t)\bigr].
\label{eq:backwardmastersum2}
\end{eqnarray}
The mean unconditional fixation time is then found via $t_{i,\sigma}=\int_0^\infty dt~t\rho_{i,\sigma}(t)$, from which we find
\begin{eqnarray}
-1 &=& 
W^+_{i,\sigma} \bigl[t_{i+1,\sigma} - t_{i,\sigma}\bigr] 
+ W^-_{i,\sigma} \bigl[t_{i-1,\sigma}- t_{i,\sigma}\bigr] \nonumber \\
&&+ \sum_{\sigma'}m_{\sigma\to\sigma'}\bigl[t_{i,\sigma'}-t_{i,\sigma}\bigr].
\label{eq:uncondcont}
\end{eqnarray}
A similar iterative equation can be found for the mean fixation time conditioned on absorption in the all-$A$ state. The only difference is the integral of $\rho_{i,\sigma}^A(t)=\partial_t\left[\sum_{\sigma'}Q_{N,\sigma';i,\sigma}(t)\right]$ is given by the fixation probability $\phi_{i,\sigma}$, and that $t_{i,\sigma}^A=\phi_{i,\sigma}^{-1}\int_0^\infty dt ~ t \rho_{i,\sigma}^A(t)$. The mean conditional fixation times, $t_{i,\sigma}^A$, therefore fulfil the relation
\begin{eqnarray}
-\phi_{i,\sigma} &=& 
W^+_{i,\sigma} \bigl[\phi_{i+1,\sigma} t_{i+1,\sigma}^A - \phi_{i,\sigma} t_{i,\sigma}^A\bigr] 
+ W^-_{i,\sigma} \bigl[\phi_{i-1,\sigma} t_{i-1,\sigma}^A- \phi_{i,\sigma} t_{i,\sigma}^A\bigr] \nonumber \\
&&+ \sum_{\sigma'}m_{\sigma\to\sigma'}\bigl[\phi_{i,\sigma'} t_{i,\sigma'}^A-\phi_{i,\sigma} t_{i,\sigma}^A\bigr].
\label{eq:condcont}
\end{eqnarray}
Structurally, equations~(\ref{eq:uncondcont}) and (\ref{eq:condcont}) are of the same form as the corresponding equations for the discrete-time model, and so they can be solved using an analogous procedure. In continuous time, however, the solution procedure no longer relies on the invertibility of the matrix $\matr{m}$ of switching rates between the states of the environment. This is because we have split up the birth-death dynamics and the changes of the environment into separate events that occur successively.

\section{Switching between two environments}\label{sec:twostate}

We now focus on the case of environments which can be in one of two possible states, i.e. $\Omega=2$. We label the two states as $\sigma=\pm1$ ($\Lambda=\{+1,-1\}$). We focus on the discrete-time scenario. The matrix $\matr{\mu}$ can then be written as
\begin{equation}
\matr{\mu}=
\begin{pmatrix}
1-p_+ & p_+ \\
p_- & 1-p_-
\end{pmatrix},
\label{eq:switchmatrix}
\end{equation}
where the quantity $p_\sigma$ ($\sigma \in \{+1,-1\}$) is the probability that the state of the environment switches to $-\sigma$ in a given time-step if it is in state $\sigma$ at the beginning of this step. We recall that our theoretical results require the inversion of $\matr{\mu}$. Excluding the case when $\Delta=\det~\matr{\mu}=1-p_+ - p_-$ vanishes, this inversion can be carried out straightforwardly,
\begin{equation}
\matr{\mu}=
\frac{1}{\Delta}
\begin{pmatrix}
1-p_- & -p_+ \\
-p_- & 1-p_+
\end{pmatrix}.
\end{equation}
For the case $\Delta=0$, we have verified that there is no anomalous behaviour of simulation results.

\subsection{Fixation probability and times}
The general result of equation~(\ref{eq:fixprobbalance4}) now reduces to the recursion
\begin{equation}
\upsilon_{i+1,\sigma} = \gamma_{i,\sigma} \upsilon_{i,\sigma} + \frac{1}{\omega^+_{i,\sigma}}\frac{p_\sigma}{\Delta}\sum_{j=1}^i(\upsilon_{j,\sigma}-\upsilon_{j,-\sigma}).
\end{equation}
The fixation probability is obtained from the set $\{v_{i,\sigma}\}$ via
\begin{equation}
\phi_{i,\sigma}=\sum_{j=1}^i\left[ \frac{1-p_{-\sigma}}{\Delta}\upsilon_{j,\sigma}-\frac{p_\sigma}{\Delta}\upsilon_{j,-\sigma}\right].\label{eq:fixprob2state}
\end{equation}
Similarly, equations~(\ref{eq:uncondbalance3}) and (\ref{eq:condbalance3}) reduce to
\begin{equation}
\nu_{i+1,\sigma} = \gamma_{i,\sigma}\nu_{i,\sigma}+\frac{1}{\omega^+_{i,\sigma}}\frac{p_\sigma}{\Delta}\sum_{j=1}^i (\nu_{j,\sigma}-\nu_{j,-\sigma}) -\frac{1}{\omega^+_{i,\sigma}}
\end{equation}
and
\begin{equation}
\eta_{i+1,\sigma} = \gamma_{i,\sigma}\eta_{i,\sigma}+\frac{1}{\omega^+_{i,\sigma}}\frac{p_\sigma}{\Delta}\sum_{j=1}^i (\eta_{j,\sigma}-\eta_{j,-\sigma}) -\frac{1}{\omega^+_{i,\sigma}}\phi_{i,\sigma}
\end{equation}
respectively. The mean unconditional and conditional fixation times are then found as
\begin{eqnarray}
t_{i,\sigma}&=& \sum_{j=1}^i\left[\frac{1-p_{-\sigma}}{\Delta}\nu_{j,\sigma}-\frac{p_\sigma}{\Delta}\nu_{j,-\sigma}\right], \\
t_{i,\sigma}^A&=& \frac{1}{\phi_{i,\sigma}}\sum_{j=1}^i\left[\frac{1-p_{-\sigma}}{\Delta}\eta_{j,\sigma}-\frac{p_\sigma}{\Delta}\eta_{j,-\sigma}\right].
\label{eq:condfixtime2state}
\end{eqnarray}

\subsection{Effective description for fast switching} \label{ssec:telegraph}
The environmental change is fast if the environmental states are short-lived, i.e. much shorter than the mean fixation time in either environment. Then we expect the population dynamics to be controlled by a set of {\em effective} transition probabilities, i.e. weighted averages of the original transition probabilities in the different environmental states. The weights are given by the fraction of time spent in each environmental state. As the dynamics of $\sigma$ follows a simple a telegraph process \cite{kampen:book:2007}, the asymptotic fraction of time spent in the state $\sigma$ is $p_{-\sigma}/(p_\sigma+p_{-\sigma})$ for $\sigma\in\{-1,+1\}$. Using this, the effective transition probabilities are given by 
\begin{equation}
\omega^\pm_{i,\eff} = \frac{p_-}{p_+ + p_-}\omega^\pm_{i,+} + \frac{p_+}{p_+ + p_-}\omega^\pm_{i,-}.
\label{eq:effectiveprobs}
\end{equation}
We note that $p_\sigma$ is the probability that in a given time-step the environment switches {\em from} state $\sigma$ to $-\sigma$. Hence the time spent in state $\sigma$ decreases with increasing $p_\sigma$ if $p_{-\sigma}$ is held fixed.

We anticipate that expression~(\ref{eq:effectiveprobs}) can formally be derived by introducing a relative scaling parameter between the switching probabilities and the birth-death probabilities, and by then taking a suitable limit in which the time scales of both processes are widely separated. We do not explore this route further here.

In this approximation the dynamics of the population are mapped to a simple birth-death process on the set $i\in\{0,1,\dots,N\}$ with absorbing states $i=0$ and $i=N$. For such processes explicit expressions for the fixation probabilities and mean fixation times are known \cite{nowak:book:2006,traulsen:bookchapter:2009,karlin:book:1981}. In the fast-switching limit we propose the following approximation for the fixation probability,
\begin{equation}
\phi_{i,\eff} = \frac{1+\sum_{k=1}^{i-1}\prod_{j=1}^k \gamma_{j,\eff}}{1+\sum_{k=1}^{N-1}\prod_{j=1}^k \gamma_{j,\eff}}.\label{eq:fixationprobtelegraph}
\end{equation}
We have here written $\gamma_{i,\eff}=\omega^-_{i,\eff}/\omega^+_{i,\eff}$.
The corresponding approximations for the mean unconditional and conditional fixation times of a single mutant are
\begin{eqnarray}
t_{1,\eff} &=& \phi_1^{\eff} \sum_{k=1}^{N-1} \sum_{l=1}^k \frac{1}{\omega^+_{l,\eff}} \prod_{m=l+1}^k \gamma_{m,\eff},\\
t_{1,\eff}^A &=& \sum_{k=1}^{N-1} \sum_{l=1}^k \frac{\phi_{l,\eff}}{\omega^+_{l,\eff}} \prod_{m=l+1}^k \gamma_{m,\eff}, \label{eq:condfixtimetelegraph}
\end{eqnarray}
respectively. These expressions \textit{exactly} describe the fixation properties of a birth-death system with the effective transition probabilities; the nature of our approximation is to assume that the birth-death process in quickly changing environments can be described by the effective transition probabilities in equation~(\ref{eq:effectiveprobs}).

Finally we note that this theory is \emph{independent} of the invertibility of the switching matrix $\matr{\mu}$.

\section{Fixation in fluctuating two-player two-strategy games}\label{sec:games}
\subsection{Evolutionary games}
As a direct application of the general theory we have developed, we now consider evolutionary game dynamics in well-mixed, finite populations. Any of the $N$ individuals can be of one of two types, $A$ or $B$. We limit the discussion to two-player games, but the extension to multi-player games (e.g. \cite{du:JRSI:2014,kurokawa:PRSB:2009,wu:Games:2013}) is straightforward.

At any point in time the environment is in one of two discrete states ($\sigma \in \{+1,-1\}$). This state fluctuates in time as specified above.  The interaction between individuals is characterised by the payoff matrix
\begin{equation}
 \begin{array}{c|cc}
  & A & B \\ \hline
  A & a_\sigma & b_\sigma \\
  B & c_\sigma & d_\sigma.
 \end{array}\label{eq:payoffmatrix}
\end{equation}
The subscript $\sigma$ indicates the dependence on the environment. The matrix focuses on the column player: A type-$A$ individual encountering another of its kind receives $a_\sigma$, and it receives $b_\sigma$ when interacting with a type-$B$ individual. In turn, an individual of type $B$ interacting with an individual of type $A$ obtains $c_\sigma$, and $d_\sigma$ is the payoff for each individual if they are both of type $B$. 

If the environment is in state $\sigma$, and if there are $i$ individuals of type $A$ in the population and $N-i$ individuals of type $B$, the expected payoffs for each type of player are
\begin{eqnarray}
\pi_A^\sigma (i) &=& \frac{i-1}{N-1}a_\sigma + \frac{N-i}{N-1}b_\sigma, \nonumber \\
\pi_B^\sigma (i) &=& \frac{i}{N-1}c_\sigma + \frac{N-i-1}{N-1}d_\sigma. \label{eq:meanpayoff}
\end{eqnarray}
The reproductive fitness of any individual is a function of the individual's payoff in the evolutionary game. We use an exponential mapping \cite{traulsen:BMB:2008,altrock:pre:2009},
\begin{eqnarray}
f_A^\sigma(i)&=&e^{\beta \pi_A^\sigma(i)}, \nonumber \\
f_B^\sigma(i)&=&e^{\beta \pi_B^\sigma(i)}.
\label{eq:fitness}
\end{eqnarray}
This is a common natural choice in which fitness is never negative and monotonically increasing with payoff. Any other functional form of the payoff-to-fitness mapping with these two properties would be equally appropriate. The constant parameter $\beta>0$ is the so-called intensity of selection. Based on this definition of fitness we model the evolutionary dynamics by the update rules of the Moran process \cite{ewens:book:2004,moran:book:1962}, which has been widely used in evolutionary game theory \cite{nowak:book:2006,traulsen:bookchapter:2009,taylor:BMB:2004}. 
The Moran process represents a simple birth-death process in which the population size remains constant, and by construction it has absorbing states at $i=0$ and $i=N$.
In a discrete-time setting, the frequency-dependent Moran process is specified by the transition probabilities \cite{nowak:nature:2004}
\begin{equation}
\omega^+_{i,\sigma} = \frac{i(N-i)}{N^2}\frac{ f_A^\sigma(i)}{\fbar^\sigma(i)}, ~~ 
\omega^-_{i,\sigma} = \frac{i(N-i)}{N^2}\frac{ f_B^\sigma(i)}{\fbar^\sigma(i)}, \label{eq:transitionprob}
\end{equation}
where $\fbar^\sigma(i)=[i f_A^\sigma(i)+(N-i) f_B^\sigma(i)]/N$ is the average fitness in the population. We note that the framework of the previous section can be applied to microscopic evolutionary dynamics other than the Moran process. This includes, for example, pairwise comparison processes \cite{blume:GEB:1993,szabo:PRE:1998}, or cases with constant selection in any one environment.

\subsection{Switching between coexistence and coordination games}
Rare mutations can introduce a previously absent strategy into the population. Typically there is only one individual of this novel type initially. We say that $B$ is the resident type, and that $A$ is the invading mutant type. All results in this section are based on the initial condition $i=1$. We chose $a_\sigma=d_\sigma=1$ for the payoff matrix. The type of game is then determined by the off-diagonal terms. We chose $b_\sigma=1+\sigma b$ and $c_\sigma=1+\sigma c$, where $b,c$ are real-valued parameters. Thus we have the payoff matrix
\begin{equation}
 \begin{array}{c|cc}
  & A & B \\ \hline
  A & 1 & 1+\sigma b \\
  B & 1+\sigma c & 1.
 \end{array}
 \label{eq:payoffmatrix2}
\end{equation}
Our parametrisation does not span the entire space of all $2\times 2$ games, but it covers some of the most common types (see below).

There exist three general types of two player-two strategy evolutionary games. First, for the coexistence game $b_\sigma>1$ and $c_\sigma>1$, selection drives the population away from the absorbing boundaries. Second, for $b_\sigma<1$ and $c_\sigma<1$, the population dynamics exhibits bi-stability. This is also known as a coordination game; selection drives the population towards the monomorphic states. In both cases there exists an internal point in frequency space for which the direction of selection changes its sign, i.e. at which the gradient of selection is zero.  This point can be calculated by solving $\omega^+_{i,\sigma}=\omega^-_{i,\sigma}$ (or equivalently ${f_A^\sigma(i)=f_B^\sigma(i)}$) for $i$, and broadly speaking it is determined by the relative magnitudes of $b$ and $c$. Third, for $b_\sigma>1, c_\sigma<1$ (or $b_\sigma<1, c_\sigma>1$) type $A$ (or type $B$) always has the higher fitness irrespective of the composition of the population. This type is then always favoured by selection, which never changes direction. 

For the remainder of this article we focus on switching between coexistence and coordination games. More precisely we choose $b>0$ and $c>0$ in (\ref{eq:payoffmatrix2}). The coexistence game corresponds to $\sigma=+1$ and the coordination game to $\sigma=-1$.

\subsection{Results}
\subsubsection*{Sample trajectory of the dynamics}
\begin{figure}[t]
\begin{center} \includegraphics[width=\textwidth]{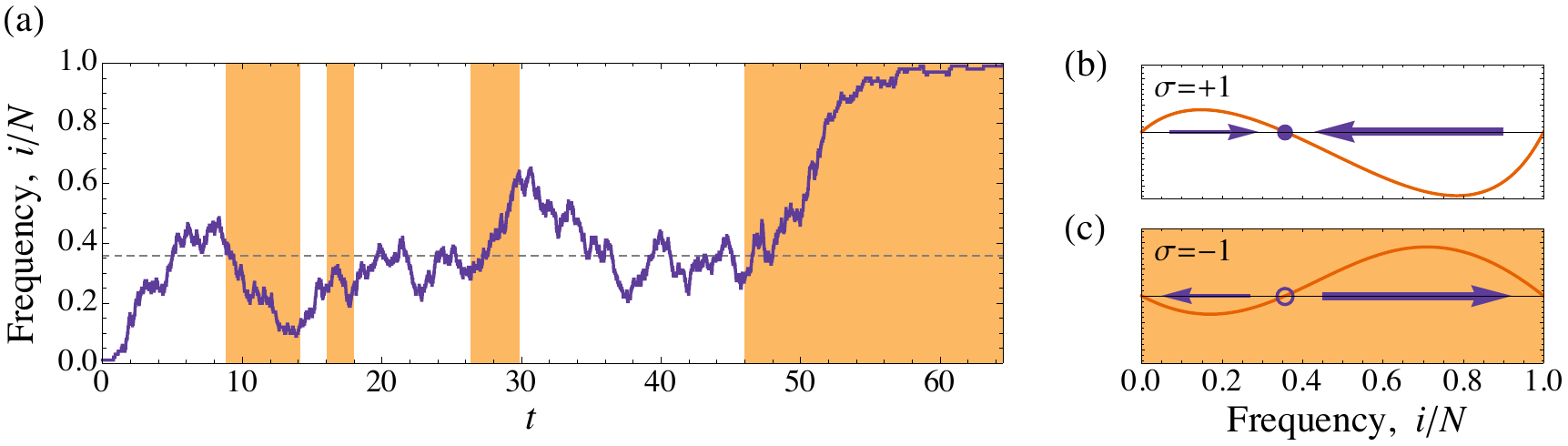}
\caption{(a) A sample trajectory (time series) of the fraction of individuals of type $A$. White background corresponds to the environment being in the $\sigma=+1$ coexistence state, while the shaded background corresponds to the $\sigma=-1$ coordination state. Dashed line is the location of the point at which selection balances, which is the same in both states of the environment.
(b) Gradient of selection in the $\sigma=+1$ coexistence state ($\omega^+_{i,+}-\omega^-_{i,+}$). Solid circle shows location of the point of selection balance, and arrows indicate the direction and magnitude of flow towards this point.
(c) Gradient of selection in the $\sigma=-1$ coordination state ($\omega^+_{i,-}-\omega^-_{i,-}$). Empty circle shows location of the point of selection balance, and arrows indicate the direction and magnitude of flow away from this point.
For the realisation in panel (a) and the selection bias shown in (b) and (c), the payoff matrix elements are $a_\sigma=d_\sigma=1$, $b_\sigma=1+0.5\sigma$ and $c_\sigma=1+0.9\sigma$, the system size is $N=100$, the selection intensity is $\beta=1$, and the switching probabilities are $p_+=10^{-3}$ and $p_-=10^{-4}$.
(Online version in colour.)}
\label{fig:fig2}
\end{center}
\end{figure}
In figure~\ref{fig:fig2}(a) we show a sample trajectory of a simulation in which a single mutant reaches fixation. The gradient of selection, $\omega^+_{i,\sigma}-\omega^-_{i,\sigma}$, for the two games is shown in figures~\ref{fig:fig2}(b) and \ref{fig:fig2}(c).  During periods when the environment is in the coexistence state (light background) the population fluctuates about the selection-balance point (dashed line), and during periods when the environment is in the coordination state (shaded background) the population is driven away from the selection-balance point. In the final period in the coordination state the mutant is driven to fixation.

\subsubsection*{Fixation probability and conditional fixation time}
\begin{figure}[t]
\begin{center} \includegraphics[width=\textwidth]{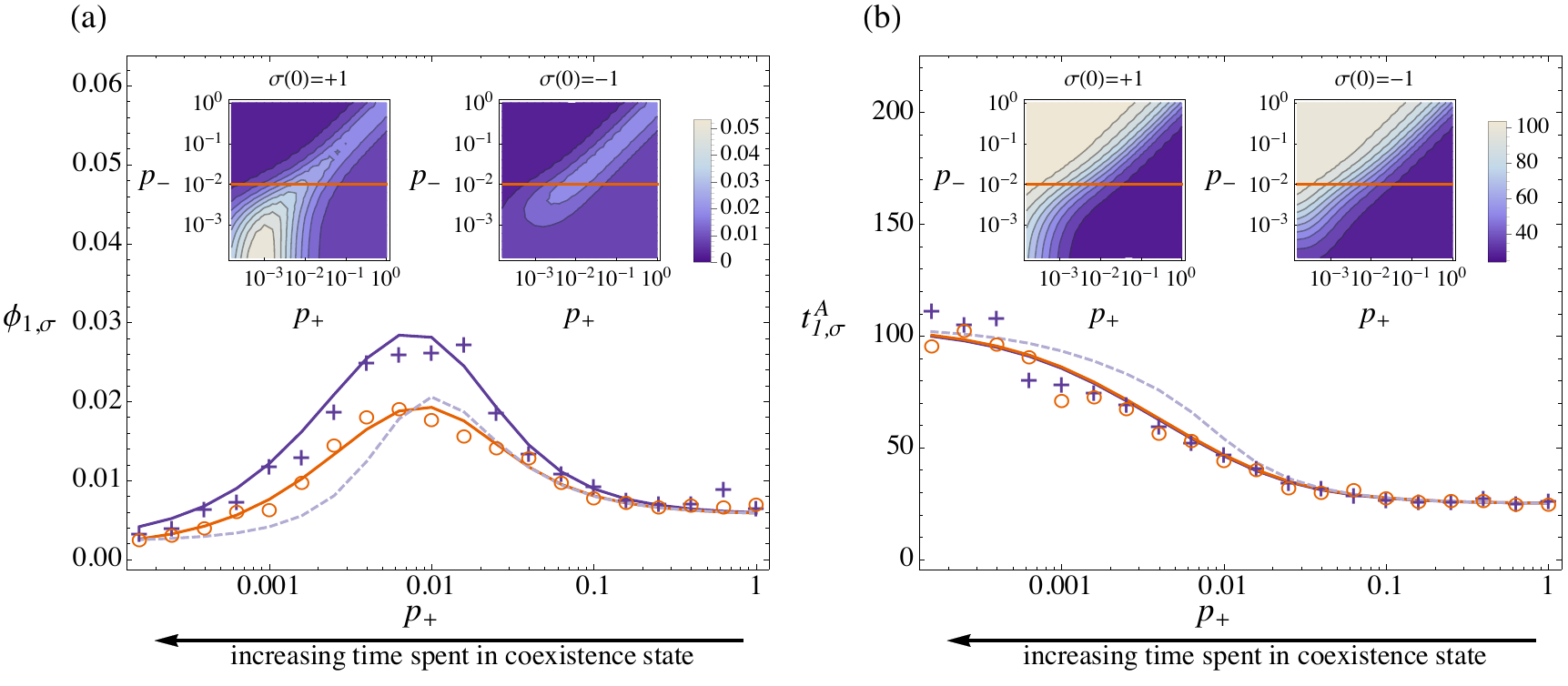}
\caption{(a) Fixation probability of a single mutant as a function of switching probabilities. 
The main panel shows simulation results (symbols; crosses correspond to $\sigma(0)=+1$ and circles to $\sigma(0)=-1$) for fixed $p_-=0.01$, along with the exact theoretical results (solid lines) from equation~(\ref{eq:fixprob2state}), and the effective theoretical result (dashed line) of equation~(\ref{eq:fixationprobtelegraph}). 
Inset panels show fixation probabilities from equation~(\ref{eq:fixprob2state}) over all combinations of $p_+$ and $p_-$. 
Left inset panel: initial condition $\sigma(0)=+1$. Right inset panel: $\sigma(0)=-1$. 
The horizontal lines correspond to the data shown in the main panel.
(b) Mean conditional fixation time (in generations) of a single mutant as a function of switching probabilities. 
The main panel shows simulation results, as described above, for fixed $p_-=0.01$, along with the exact theoretical results (solid lines) of equation~(\ref{eq:condfixtime2state}), and the effective theoretical result (dashed line) of equation~(\ref{eq:condfixtimetelegraph}). 
Inset panels show mean conditional fixation times from equation~(\ref{eq:condfixtime2state}) over all combinations of $p_+$ and $p_-$. 
Left inset panel: initial condition $\sigma(0)=+1$. Right inset panel: initial condition $\sigma(0)=-1$. 
The horizontal lines correspond to the data shown in the main panel.
The payoff matrix elements are $a_\sigma=d_\sigma=1$, $b_\sigma=1+0.5\sigma$ and $c_\sigma=1+0.9\sigma$, the system size is $N=50$, and the selection intensity is $\beta=0.5$.
(Online version in colour.)}
\label{fig:fig3}
\end{center} 
\end{figure}
In figure~\ref{fig:fig3} we show the effect of switching the environment on the fixation dynamics. We choose $c>b>0$. By equating the reaction probabilities (i.e., setting $\omega_{i,\sigma}^+=\omega_{i,\sigma}^-$), or equivalently equating the expected payoffs in equation (\ref{eq:meanpayoff}), and looking at leading order terms in $N$, the gradient of selection is seen to change sign at $i^*/N \approx b/(b+c)<1/2$. This point is closer to the extinction state of the mutant ($i=0$) than to the fixation state ($i=N$). We next describe the key observations we make from these results, before we turn to their interpretation.

\emph{Fixation probability (figure~\ref{fig:fig3}(a)):}\\
The fixation probability in this example depends non-trivially on the rates with which the environment switches states; we find an optimal combination of switching rates, $p_+\simeq p_-$, for which fixation of a single mutant is most likely. The fixation probability is dependent on the initial state of the environment for $p_\sigma \lesssim 0.1$.

\emph{Fixation time (figure~\ref{fig:fig3}(b)):}\\ 
Mean conditional fixation times show very little dependence on the initial state of the environment. The fixation time is small for $p_+>p_-$ when the environment is found mostly in the coordination game, and large when the environment is mostly in the coexistence state.

\emph{Validity of the theoretical approach:}\\
As seen in both panels of figure~\ref{fig:fig3} the theoretical predictions of equations (\ref{eq:fixprob2state}) and (\ref{eq:condfixtime2state}), indicated by solid lines, are in convincing agreement with simulation data. Theoretical results from the model with effective transition rates (section~\ref{ssec:telegraph}) reproduce the simulation data qualitatively. Quantitative agreement is obtained in the limit of large switching rates, but unsurprisingly, there are systematic deviations when switching is slow.

\subsubsection*{Interpretation}
We now proceed and give an intuitive explanation for the observed effects.

\emph{Mean conditional fixation time is reduced as more time is spent in coordination environment:}\\
The behaviour of the fixation time can intuitively be understood from the deterministic gradient of selection of the two games (figure~\ref{fig:fig2}(b) and \ref{fig:fig2}(c)).
If fixation happens, it will generally be quicker in the coordination game than in the coexistence game \cite{antal:BMB:2006,altrock:njp:2009}. This is due to the adverse selection bias in the coordination game at low mutant numbers (see figure~\ref{fig:fig2}(c)). The more time the system spends in this region of adverse selection the less likely it is for the mutant to reach fixation. Thus if fixation happens in a coordination game then it happens fast.
In the coexistence game on the other hand the direction of selection is towards the balance point, as shown in figure~\ref{fig:fig2}(b). The system can `afford' to spend significant time in the region of small mutant numbers and still reach fixation eventually even after repeated excursions throughout frequency space. There is thus no need for fixation to occur quickly, and conditional fixation times can be long.

These observations make it plausible that the mean conditional fixation time will generally decrease when less time is spent in the coexistence game, which is exactly what we find in figure~\ref{fig:fig3}(b). We have tested other choices of the parameters $b$ and $c$ for which the two games are a coexistence game and a coordination game, and we find that the behaviour of the mean conditional fixation times is robust under these changes.

\emph{Mean conditional fixation time is largely independent of the initial state of the environment:}\\
Systems started in the coordination environment will tend to reach extinction relatively quickly due to initial adverse selection, unless the environment switches to the coexistence state early on. Thus the sample of runs that reach fixation started from the coordination-game environment will be dominated by runs in which the environment switches soon after the start of the run. 
Then we expect that the value of the mean conditional fixation time is close to the one obtained when starting in the coexistence game.

\emph{Dependence of fixation probability on initial state of the environment:}\\
The data in figure~\ref{fig:fig3}(a) show that initiating the dynamics in the coexistence game favours fixation of the mutants for $p_+ \lesssim 0.1$ (when $p_-=0.01$ is fixed), however above this threshold the initial state of the environment has relatively little effect. The reason for this is as follows. When starting in the coordination game, selection pushes the mutant towards extinction. Hence fixation is more likely if the initial state is the coexistence fitness landscape. Above $p_+\simeq 0.1$, the switching process of the environment is too fast for the initial condition to have any significant effect on the population dynamics. It is this regime in which we expect the effective description (section~\ref{ssec:telegraph}) to approximate the system well. This is indeed confirmed in figure~\ref{fig:fig3}, the theoretical prediction of the effective theory, equation~(\ref{eq:fixationprobtelegraph}), agrees well with our simulation results in this fast-switching region.

\begin{figure}[t]
\begin{center} \includegraphics[width=\textwidth]{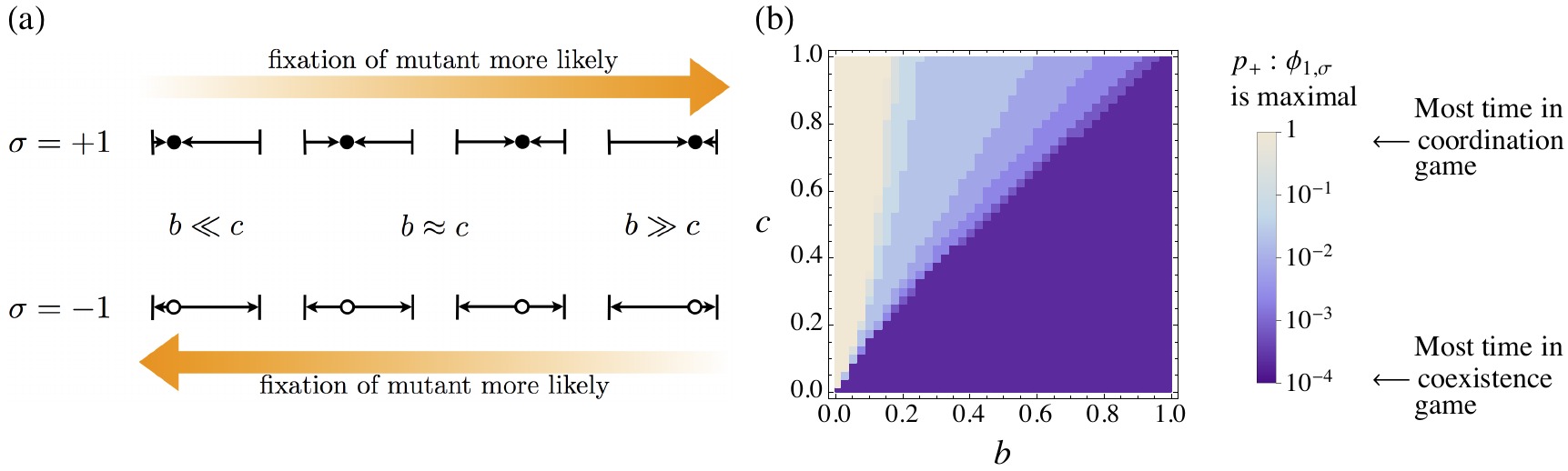}
\caption{(a) Illustration of selection bias in the two environments for different locations of the balance point; (b) Value of $p_+$ at which $\phi_{1,\sigma}$ is maximal given $p_-=0.01$ as a function of $b$ and $c$. Remaining parameters are $\beta=0.5$ and $N=50$.
(Online version in colour.)
}
\label{fig:fig4}
\end{center} 
\end{figure}

\emph{Behaviour of fixation probability depends on location of the selection-balance point:}\\
If the environment is fixed to the coexistence-game state, fixation is more likely the closer the point of selection balance is to the fixated state, see figure~\ref{fig:fig4}(a). The location of this balance point is approximated by $i^*/N=1/(1+c/b)$, and so the fixation probability increases as $c/b$ is decreased. In a fixed coordination-game environment the reverse is the case. The range of adverse selection is to the left of the balance point, and so fixation is less likely the closer the point of selection balance is to the fixated state.

For $b\ll c$, i.e. a selection-bias point close to $i=0$, we therefore expect that the fixation probability will increase the more time is spent in the coordination-game environment, i.e. $\phi_{1,\sigma}$ is an increasing function of the probability $p_+$ with which the system leaves the $\sigma=+1$ state (coexistence game). This is indeed what we find in simulations (data not shown). For $b\gg c$, i.e. $i^*$ close to $i=N$, the reverse is the case. Fixation is more likely in the coexistence game ($\sigma=+1$), and the fixation probability is hence a decreasing function of $p_+$ at fixed $p_-$. Although we do not show the data here, this is again confirmed in simulations.

For $b\approx c$ the situation is less clear. The fixation probability will be comparable in both games if the environment is frozen. Two effects here conspire to produce a non-trivial outcome:
\begin{itemize}
\item[(i)] Consider the case in which the system is mostly in the coordination-game state, i.e. $p_+\gg p_-$. It is plausible that an occasional switch to a coexistence game will make fixation more likely than in a constant coordination game. This is because the coexistence-game environment pushes the system away from extinction at low mutant numbers. In the regime of $p_+\gg p_-$ we thus expect the fixation probability to increase as $p_+$ is lowered. In other words, $\phi_{1,\sigma}(p_+)$ is a decreasing function at large $p_+$.
\item[(ii)] Similarly, if the system is mostly in the coexistence-game environment ($p_+\ll p_-$), short periods of time in the coordination game can make fixation more likely. This is because selection at large mutant numbers is directed  towards fixation in the coordination game. At $p_+\ll p_-$ we expect $\phi_{1,\sigma}$ to be an increasing function of $p_+$. 
\end{itemize}
These two effects taken together generate a maximum of the fixation probability at intermediate values of $p_+\approx p_-$, which is exactly what we find in figure~\ref{fig:fig3}(a). We would like to stress that the effect (i) is only present provided the selection-balance point is not too close to the extinct state. The phenomenon discussed under (ii) is only present if the selection-balance point is not too close to the fixated state. If the balance point is located too close to either boundary the corresponding effect will be suppressed and the remaining effect dominates. One then finds monotonically increasing or decreasing dependences $\phi_{1,\sigma}(p_+)$.

To confirm our picture we varied the payoff parameters $b$ and $c$, and find the value of $p_+$ that maximises fixation probability for a given $p_-=0.01$, as a function of $b$ and $c$ in figure~\ref{fig:fig4}(b). The point of selection balance is approximately $1/(1+c/b)$ (up to system-size corrections). The presence of diagonal structures in figure~\ref{fig:fig4}(b) shows that the behaviour of the fixation probability is dependent on the location of the selection-balance point. If this point is close to the fixation state $i=N$ ($b\gg c$, bottom-right in figure~\ref{fig:fig4}(b)), then the fixation probability is maximal for vanishing $p_+$. If this point is close to the extinction state ($b\ll c$, top-left in figure~\ref{fig:fig4}(b)), then the fixation probability is maximal for large $p_+$. For intermediate locations of the selection balance point ($b\approx c$) fixation is maximised at a non-trivial combination of environment states. 

The features observed in figure~\ref{fig:fig3}, i.e. the peak in the fixation probability and shape of the mean conditional fixation time as a function of $p_+$, are found to be robust when the system size is increased. Fixation probabilities generally decrease with system size but the observed peak becomes sharper. The mean conditional fixation times in a frozen coexistence environment scale exponentially with $N$, whereas they scale as the logarithm of $N$ in the coordination environment \cite{mobilia:JTB:2010}. We observe these scalings in our system, with the mean conditional fixation time increasing exponentially with $N$ for small $p_+$, and increasing sub-linearly with $N$ for large $p_+$.

\section{Mutation-selection equilibria under fluctuating environments}\label{sec:mut}
\subsection{Mutations and stationary distributions}
We now consider systems with mutations occurring during the dynamics. This removes the possibility of fixation and extinction. The combination of mutation, selection, and noise can lead to non-trivial stationary states. We introduce mutation by modifying the discrete-time transition probabilities of equation~(\ref{eq:transitionprob}) and now use
\begin{equation}
\hat{\omega}^+_{i,\sigma} = (1-u)\frac{i(N-i)}{N^2}\frac{ f_A^\sigma(i)}{\fbar^\sigma(i)} + u\frac{(N-i)^2}{N^2}, ~~ 
\hat{\omega}^-_{i,\sigma} = (1-u)\frac{i(N-i)}{N^2}\frac{ f_B^\sigma(i)}{\fbar^\sigma(i)} + u\frac{i^2}{N^2},
\label{eq:transitionprobmut}
\end{equation}
where $u\ll1$ is the mutation rate. The transition probabilities $\hat{\omega}^+_{0,\sigma} = \hat{\omega}^-_{N,\sigma} = u$ are now non-zero, and so the states $i=0$ and $i=N$ are no longer absorbing.

The stationary probability $\rho_{i,\sigma}$ of finding the system in state $(i,\sigma)$ ($i=0,1,\dots,N$, $\sigma\in\Lambda$) is obtained as the solution of the balance equation
\begin{equation}
\rho_{i,\sigma} = \sum_{\sigma'\in\Lambda} \mu_{\sigma'\to\sigma}\left[\hat{\omega}^+_{i-1,\sigma'}\rho_{i-1,\sigma'}+\hat{\omega}^-_{i+1,\sigma'}\rho_{i+1,\sigma'} +(1-\hat{\omega}^+_{i,\sigma'}-\hat{\omega}^+_{i,\sigma'})\rho_{i,\sigma'}\right].
\label{eq:statdist}
\end{equation}
This equation is of the form $\rho_{i,\sigma}=\sum_{\sigma'} \sum_j \hat{R}_{(j,\sigma') \to (i,\sigma)} \rho_{j,\sigma'}$, and it is solved by finding the eigenvector corresponding to the eigenvalue $\lambda=1$ of the linear operator $\hat R$. In principle this can be done analytically, but we use standard numerical packages to find the eigenvector. The stationary distribution for the state of the population is found by summing over all states of the environment, $\rho_i=\sum_\sigma\rho_{i,\sigma}$. This solution is exact.

If the environment states are long-lived, the population will relax to the stationary state of the current environment before the next switching event. With this, one might expect that the overall stationary distribution is the weighted average of the stationary distributions one would obtain in the respective stationary environments. The stationary distribution in a single fixed environment, $\sigma$, can be found explicitly as
\begin{equation}
\rho^{(1)}_{i,\sigma} = \Gamma_{i,\sigma} \rho^{(1)}_{0,\sigma}, ~~
\Gamma_{i,\sigma} = \prod_{j=1}^i \frac{\hat{\omega}^+_{j-1,\sigma}}{\hat{\omega}^-_{j,\sigma}}, ~~
\rho^{(1)}_{0,\sigma} = \left(1+\sum_{i=1}^N\Gamma_{i,\sigma}\right)^{-1}.
\label{eq:statdistsingle}
\end{equation} 
This can be derived for example from equation~(\ref{eq:statdist}) assuming that the transition matrix of the environment is diagonal, $\mu_{\sigma\to\sigma'}=\delta_{\sigma\sigma'}$, where $\delta_{\sigma\sigma'}$ is the Kronecker delta. The average stationary distribution over many slow-switching environments can then be written as
\begin{equation}
\overline{\rho}_i = \sum_{\sigma'\in\Lambda} \rho_{\sigma'} \rho^{(1)}_{i,\sigma'},
\label{eq:statdistave}
\end{equation}
where $\rho_{\sigma}$ is the probability that the environment is in state $\sigma$.

Alternatively, if the switching probabilities per time-step are large one might expect the stationary distribution to be approximated by the distribution found in a system controlled by the effective transition rates, $\hat{\omega}^\pm_{i,\eff}$. These are obtained as described in section~\ref{ssec:telegraph}, with suitable modifications to account for mutation. The resulting stationary distribution is found as
\begin{equation}
\rho_{i,\eff} = \Gamma_{i,\eff} \rho_{0,\eff}, ~~
\Gamma_{i,\eff} = \prod_{j=1}^i \frac{\hat{\omega}^+_{j-1,\eff}}{\hat{\omega}^-_{j,\eff}}, ~~
\rho_{0,\eff} = \left(1+\sum_{i=1}^N\Gamma_{i,\eff}\right)^{-1}. \label{eq:statdisteff}
\end{equation}

The distributions $\overline{\rho}_i$ and $\rho_{i,\eff}$ are both approximations. To evaluate the validity of the assumptions leading to these approximations we compute the distance
\begin{equation}
 d(t)=\frac{1}{2}\sum_{i=0}^N \left| \overline{\rho_i} - P_i^{\rm sim}(t) \right|, \label{eq:distance}
\end{equation}
and similarly for $\rho_{i,\eff}$, where $P_i^{\rm sim}(t)$ is distribution of the population at time $t$ obtained from simulations. To confirm our analytical approach we also compute the distance of simulation data from the exact solution for $\rho_i$. We allow the system to run for a fixed time $T$, and we then use the time-averaged distance, $\overline{d}=(2/T)\int_{T/2}^T d(t)\;dt$ to evaluate the accuracy of the approximations. We ignore the first half of the time series to remove remnants of the initial condition.
We note that the time $T$, measured in generations, is equivalent to $NT$ simulation time-steps and is chosen to be long enough such that the system relaxes to the stationary state before measurements start at time $T/2$.

\subsection{Results}
\begin{figure}[t]
\begin{center} \includegraphics[width=\textwidth]{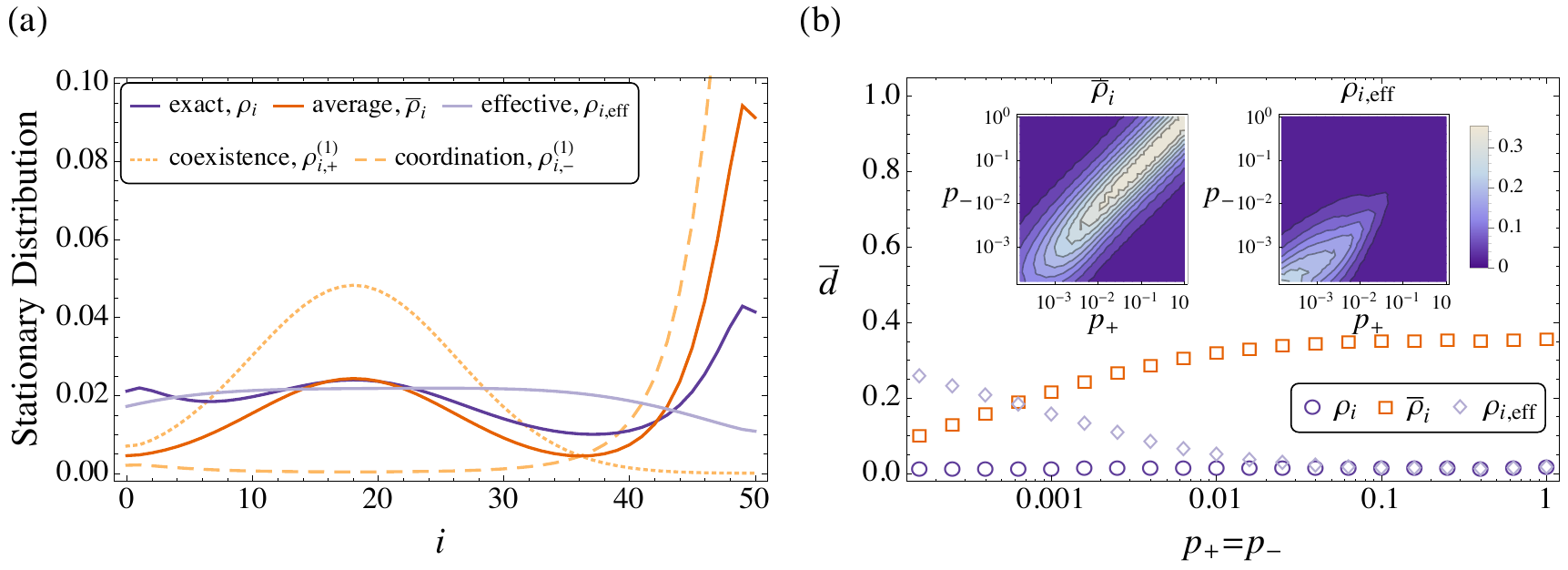}
\caption{(a) The stationary distributions in the single-environment coexistence game $\rho_{i,+}^{(1)}$ (dotted line) and coordination game $\rho_{i,-}^{(1)}$ (dashed lines) calculated from equation~(\ref{eq:statdistsingle}), along with the exact solution $\rho_i$ [equation~(\ref{eq:statdist}), evaluated numerically] and the `average' $\overline{\rho}_i$ [equation~(\ref{eq:statdistave})] and `effective' $\rho_{i,\eff}$ [equation~(\ref{eq:statdisteff})] approximate stationary solutions (solid lines).
These distributions are for switching probabilities $p_+=p_-=10^{-3}$.
(b) The time-averaged distance [equation~(\ref{eq:distance})] between distributions obtained from an ensemble of $5\times10^5$ simulations, $Q_i(t)$, and the analytic stationary distributions for symmetric switching, $p_+=p_-$. Simulations run for $T=2\times10^3$ generations. Inset plots show the time-averaged distances over all switching-parameter space. Left inset panel shows the distance to the `averaged' stationary solution [equation~(\ref{eq:statdistave})], and right inset panel shows the distance to the `effective' stationary solution [equation~(\ref{eq:statdisteff})].
The payoff matrix elements are $a_\sigma=d_\sigma=1$, $b_\sigma=1+0.5\sigma$ and $c_\sigma=1+0.9\sigma$, the system size is $N=50$, the selection intensity is $\beta=0.5$, and the mutation probability is $u=0.02$.
(Online version in colour.)
}
\label{fig:fig5}
\end{center} \end{figure}

We present results for the two-world scenario, where the environment switches between a coexistence game and a coordination game as described above. The stationary distribution of the environmental state is given by $\rho_{\sigma=+1}=p_-/(p_+ + p_-)$ and $\rho_{\sigma=-1}=p_+/(p_+ + p_-)$.

The stationary distributions of the population for the fixed environments (calculated using equation~(\ref{eq:statdistsingle})) are shown in figure~\ref{fig:fig5}(a). In a constant coexistence game ($\sigma=+1$) the stationary distribution is peaked about the point at which the gradient of selection changes sign, and in a fixed coordination-game environment ($\sigma=-1$) we find a distribution which is strongly peaked near the $i=N$ state. The asymmetry is due to the imbalanced payoff matrix used, such that the basin of attraction for the $i\simeq N$ state is much larger than for the $i\simeq 0$ state. For the parameters chosen in the figure \ref{fig:fig5}, the selection-balance point is at $i^*\approx 18$.

For equal switching rates, $p_+=p_-$, the averaged stationary distribution $\overline{\rho}_i$ lies exactly in between the two single-environment distributions. The effective distribution is approximately uniform in the centre of the domain, with a lower probability of being found close to the domain boundaries. This reflects the fact that for equal switching probabilities the effective game is close to neutral, but frequent mutations push the population to the interior. The exact solution [equation~(\ref{eq:statdist})] matches the features of the single-environment distributions, with a peak at $i\simeq N$ and at the coexistence point $i^*$. Interestingly, this solution also predicts a peak at the $i\simeq 0$ state, a feature which is not seen in the single-environment distributions, or in the effective distribution.

As seen in the main panel of figure~\ref{fig:fig5}(b), the exact solution is confirmed by simulations across many orders of magnitude of switching probabilities. Any deviations can be attributed to incomplete equilibration of the measure used in equation~(\ref{eq:distance}). For large switching probabilities, the effective stationary distribution, $\rho_{i,\eff}$, matches the simulations. As expected the effective theory becomes inaccurate for slow switching, roughly below $p_\sigma\simeq 10^{-2}$, in our example. The weighted-average stationary distribution, $\overline{\rho}_i$, shows the opposite behaviour. It is in reasonable agreement with simulations for slow switching, but shows systematic deviations when the switching process is too fast for the population to react adiabatically.

This picture is further corroborated by the data shown in the inset panels of figure~\ref{fig:fig5}(b). The weighted-average and the effective distributions accurately predict the stationary distribution obtained from simulations when the two switching rates are very disparate, i.e. $p_+\ll p_-$ or vice versa (top-left and bottom-right corners of the two insets). In these regions, the environment spends most of its time in one state, so that the model effectively reduces to the single-environment case. Simulation data, the exact solution, and both approximations then all collapse to the same result, the stationary distribution obtained in a single fixed environment.

The approach based on effective transition rates (right inset of figure~\ref{fig:fig5}(b)) is found to be accurate over a large range of switching probabilities away from the slow-switching scenario. Conversely, the weighted-average distribution (left inset of figure~\ref{fig:fig5}(b)) becomes increasingly accurate if the dynamics of the environment is slow ($p_\sigma\to0$).

\section{Summary and conclusion}\label{sec:sum}
The dynamics of a population evolving under changing environmental conditions is an important concept in the study of bacterial populations. Previous work has focused on deterministic analyses \cite{kussell:Genetics:2005}, or on an environment following a continuous stochastic process \cite{assaf:PRL:2013}. Here we have taken a different route and assumed that the environment switches between discrete states. We have developed the mathematical formalism to describe fixation properties in a general birth-death process in an environment fluctuating between an arbitrary number of states. The main results of this investigation are self-consistent expressions for the fixation probability of a mutant in a fixed-size population, as well as for the mean unconditional and conditional fixation times.  For short-lived environments we put forward an approximation based on effective transition probabilities. 

As a specific application we discuss the fixation properties in the context of an evolutionary game in a two-world scenario. The two states of the environment then correspond to two different payoff matrices of the underlying games. Simulations confirm our exact solution over a wide range of switching probabilities. The approximation based on effective transition probabilities is seen to reproduce simulation data in the limit of fast switching.

Focussing on the case of switching between a coexistence game and a coordination game we find unexpected non-trivial behaviour of the fixation probability of a single mutant. We observe in our analytical results and in simulations that fixation can be more likely in a scenario in which the fitness landscape switches between the two games than in either of the two constant environments. We provide an intuitive explanation for this effect, and we have investigated in detail the circumstances under which this phenomenon can occur.

Adding mutations to the dynamics removes the possibility of fixation, but introduces non-trivial stationary states. We develop a method for the calculation of this distribution, along with approximations for long-lived and short-lived environmental states respectively. These approximations are shown to agree well with simulations in their respective limits. 

The general theory developed here now allows further investigation of evolutionary dynamics in time-varying environments. It provides a first mathematical characterisation of the effects one may expect in such systems. The closed form self-consistent solutions will help to speed up future studies, and they may remove the need for extensive computer simulations.

While our work is mainly mathematical, we think that our theory can be used to interpret existing experimental studies such as those studied by Acar {\em et al.~}\cite{acar:NG:2008}. For some biological systems it may be more appropriate to use constant selection in each environment, as opposed to frequency-dependent selection. Our example of switching between coexistence and coordination games was chosen to illustrate the theory. We have also seen such cases lead to unexpected effects. We note that both types of game have been observed in systems of experimental evolution \cite{maclean:PloSB:2010,gore:Nature:2009,maclean:Nature:2006}. We hope the formalism we have developed will be useful to analyse models closer to other biological applications, and potentially to guide future experiments on evolutionary systems in time-dependent environments.

On a more general level constructing a mathematical theory of evolutionary dynamics is very much work in progress. An integral part of the evolution of microbes and higher organisms alike is frequency-dependent selection. At the same time external factors determining the detailed mechanics of selection may vary in time. In this work we have combined frequency-dependent selection, fluctuating environments, and stochastic dynamics in finite populations into one model, and we have provided the analytical tools for its analysis. This, we hope, is a contribution toward a more complete understanding of evolutionary processes.

\subsection*{Acknowledgements}
P.A.~acknowledges support form the EPSRC. P.M.A.~gratefully acknowledges financial support from Deutsche Akademie der Naturforscher Leopoldina, grant no.~LPDS 2012-12, as well as discussions with Whan Ghang and Christian Hilbe.

\end{document}